\documentclass[pra,aps,twocolumn,superscriptaddress,showpacs]{revtex4-1}
\usepackage{CJK}
\usepackage{graphicx}
\usepackage{dcolumn}
\usepackage{subfigure}
\usepackage{bm}
\usepackage{amsmath}
\usepackage{amssymb}
\usepackage{color,soul}
\usepackage{xcolor}		
\usepackage{bbold}
\usepackage{mathptmx}
\usepackage[pagebackref=false,colorlinks,allcolors=blue,citecolor=blue]{hyperref}

\DeclareMathAlphabet{\mathcal}{OMS}{cmsy}{m}{n}

\begin{document}
\begin{CJK*}{GB}{}

\title{Fast and efficient deterministic quantum state transfer between two remote mechanical resonators}
\author{Mojtaba Rezaei}
\email{m.rezaei@mail.um.ac.ir}
\affiliation{Department of Physics, Ferdowsi University of Mashhad, Mashhad, PO Box 91775-1436, Iran}
\author{Kurosh Javidan}
\email{javidan@um.ac.ir}
\affiliation{Department of Physics, Ferdowsi University of Mashhad, Mashhad, PO Box 91775-1436, Iran}
\author{Hamidreza Ramezani}
\email{hamidreza.ramezani@utrgv.edu}
\affiliation{Department of Physics and Astronomy, University of Texas Rio Grande Valley, Edinburg, TX 78539, USA}
\author{Mehdi Abdi}
\email{mehabdi@iut.ac.ir}
\affiliation{Department of Physics, Isfahan University of Technology, Isfahan 84156-83111, Iran}
\date{\today}

\begin{abstract}
The main challenge in deterministic quantum state transfer in long-distance quantum communications is the transmission losses in the communication channel. To overcome this limitation, here we use the adiabatic theorem and find a lossless evolution path between two remote mechanical modes. By adiabatic variation of the effective coupling strengths between the two nodes and the intermediate optical channel modes, we engineer a transmission path for the quantum state transfer that is decoupled from the decaying fiber modes. Using our proposed method we show that one obtains a quantum state transfer with high efficiency. Furthermore, to bypass the slow nature of the adiabatic process and its sensitivity to the mechanical damping and noise as well as the strength of the driving pulses, we develop the shortcut to adiabatic passage protocol for our proposed quantum state transfer. Our results show that the shortcut to adiabaticity provides an efficient and fast quantum state transfer even for small values of the coupling strength. We show that the performance of our protocol for long-distance quantum communications remains efficient for transferring the quantum states between two remote mechanical resonators being hundred meters apart.
\end{abstract}
\maketitle
\end{CJK*}
\section{\label{sec1}introduction}

The ability of quantum state transfer (QST) with high efficiency between two nodes in a quantum network is a key task for the realization of quantum communications \cite{kimble2008quantum,northup2014quantum,hammerer2010quantum,reiserer2015cavity}. A quantum network consists of nodes, which are clusters of stationary quantum memories, connected by quantum communication channels, such as waveguides. The quantum information in quantum bits (qubits) are stored in the nodes and are transmitted across the desired distance through quantum channel by means of photons or phonons as flying qubits \cite{muller1996quantum,ursin2007entanglement,ritter2012elementary,tiecke2014nanophotonic,goban2015superradiance}.

Among the others, there are two well-known protocols in performing deterministic QST between remote nodes of a quantum network. The standard protocol which is based on wave packet shaping (WPS), employs special laser drive which excites the sending node such that its state is mapped into the flying qubits that will be perfectly absorbed by the receiving node~\cite{cirac1997quantum,ritter2012elementary,stannigel2011optomechanical}. The photon losses in the quantum channel and the issue of temporal-envelop mismatch in the absorption and emission processes are the dominant challenges for the deterministic QST in the standard protocol~\cite{habraken2012continuous}. Another protocol for QST is based on the stimulated Raman adiabatic passage which has been developed to surpass the  above mentioned problems~\cite{kovachy2012adiabatic,bergmann1998coherent,moller2008quantum,klein2007robust,du2014experimental,lacour2006implementation}. During the adiabatic passage (AP) protocol, the quantum states are preserved in a dark state that decouples from the channel dissipation. For realizing the AP protocol, the classical driving pulses of both nodes are applied in a counterintuitive order in which the classical driving field in the receiving node is switched on prior to the one in the sending node \cite{vitanov1997analytic,pellizzari1997t}. 

In recent years, optomechanical systems have offered an excellent approach for implementing the QST protocols between mechanical modes as stationary nodes \cite{hill2012coherent,liu2013electromagnetically,bochmann2013nanomechanical,andrews2014bidirectional,shkarin2014optically,andrews2015quantum,lecocq2016mechanically,balram2016coherent}. In optomechanical systems, a mechanical mode can couple to any of the optical modes of a cavity via radiation pressure force \cite{aspelmeyer2014cavity}. Hence, one mechanical resonator can act as a mediator
 for the QST between two  optical modes \cite{stannigel2010optomechanical,safavi2011proposal}. Also, one can provide a three-mode system in which two mechanical modes couple to a common optical mode to realize the QST between the mechanical resonators based on the WPS approach \cite{xu2016topological,ockeloen2016quantum,dong2014optomechanically,pontin2016dynamical,shkarin2014optically,massel2012multimode,spethmann2016cavity,dong2014optomechanically,felicetti2017quantum}.

Here, we investigate the efficient methods for performing the QST between two remote mechanical modes inside optomechanical systems that are connected via an optical fiber. More specifically, we offer a method for transferring the quantum states among mechanical resonators that is not sensitive to transmission losses in the communication channel.
We employ the AP protocol for the QST in which a quantum state is preserved in a dark mode, which is a linear combination of mechanical resonator modes, with negligible excitation to the optical channel modes. As a result, a high efficiency QST protocol between the two mechanical modes is attained. An advantage of our proposed method  with respect to WPS protocol is that we do not require the exact control of the temporal shape of the driving pulses of both nodes during the transfer process.

Generally, the adiabatic process is achieved in the limit of large transfer operation times or large driving pulses strength. Therefore, in the AP protocol the QST is slow and can suffer from dissipations in the sending and receiving nodes that create decoherence in the entire transfer process.
Also, due to the requirement on adiabaticity in the AP protocol, the transfer efficiency is sensitive to the driving pulses strength. To overcome these restrictions, we therefore employ the shortcut to adiabatic passage (STAP) approach in the optomechanical system to speed up the state transfer between the two mechanical resonators. In this protocol, according to the transitionless quantum driving algorithm \cite{berry2009transitionless,chen2010shortcut,chen2011lewis,giannelli2014three}, the diabatic transitions among the adiabatic eigenmodes are suppressed by adding auxiliary counter-diabatic processes. This leads to a fast and perfect state transfer through the dark mode evolution.
In contrast to the AP protocol which has a challenge in conflicting between transfer speed and efficiency \cite{PhysRevLett.122.050404}, the QST becomes perfect for short operation times even with small values of the coupling strength in the STAP protocol. As a result, the  transfer efficiency becomes very close to unity for a wide range of coupling strengths.

The paper is presented in four sections. In Sec. \ref{sec2} we introduce the model and protocols. Section \ref{sec3} is devoted to the results and discussions, while the work is summarized in Sec. \ref{sec4}.

\section{\label{sec2}model and protocols}

\subsection{\label{sec21}The model}
\begin{figure}[b]
	\includegraphics[width=\columnwidth]{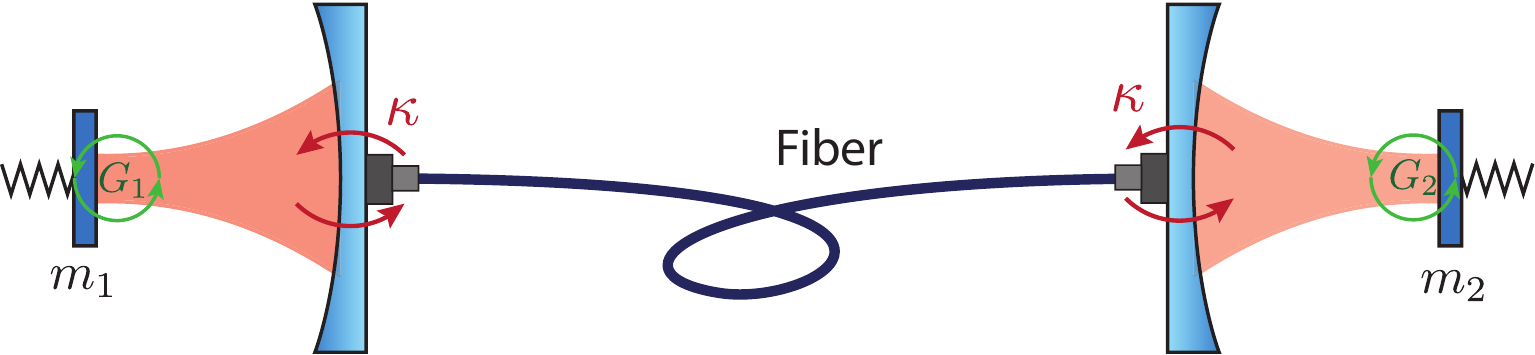}
	\caption{Schematic setup of the basic quantum network for the realization of the deterministic QST between two nodes. Nodes are optomechanical systems that are driven by laser and connected by an optical fiber. Each node has a mechanical resonator. The mechanical  modes $m_{1}$ and $m_{2}$ are coupled to the fiber modes $f_{n}$ with effective strengths $g_{1}(t)$ and $g_{2}(t)$, respectively. The $\gamma_{\rm fib}$ is the loss rate of the optical fiber. \label{diag1}}
\end{figure}

Our system is composed of two similar nodes each containing an optomechanical system that are connected by an optical fiber as shown in Fig.~\ref{diag1}.
The system Hamiltonian is
\begin{equation}
H=H_{\rm OM}+H_{\rm fib}+H_{\rm int}.
\end{equation}
Above the local dynamics of the nodes is given by the optomechanical Hamiltonian $H_{\rm OM}$. Furthermore, the optical fiber modes Hamiltonian in a frame rotating with the driving laser frequency ($\omega_{l}$) reads as
\begin{equation}
H_{\rm fib} =\hbar\sum_{n=-\infty}^{+\infty}\Delta_{n} f_{n}^{\dagger}f_{n}=\hbar\!\sum_{n= -\infty}^{+\infty}(\Delta_{0}+n\delta_{\textsc{fsr}}) f_{n}^{\dagger }f_{n},
\end{equation}
where $\Delta _{n}=\omega _{n}-\omega _{l}$ is the detuning. Here, $\omega _{n} =\omega_{0} +n\delta_{\textsc{fsr}} $ and $f_{n} $ are the optical frequency and the annihilation operator of $n$th mode of the fiber, respectively.  $\omega_{0}$ is frequency of the fiber mode $f_0$ that has the closest frequency to the laser drive. The fiber modes are separated by a free spectral range $\delta_{\textsc{fsr}}=\pi c /L$ (with $c$ being the speed of light in fiber).

The optical fiber modes are coupled to the mechanical resonators through the cavity modes that in turn interact with the mechanical modes via radiation pressure.
After adiabatic elimination of the cavity modes we arrive at the following effective interaction Hamiltonian between the fiber modes and the mechanical resonators
\begin{equation}\label{H cav-fib}
H_{\rm int}= \hbar\!\sum_{n=-\infty}^{+\infty}\left[g_{1}(t)m_{1}^{\dag } +(-1)^{n} g_{2}(t) m_{2}^{\dag } \right]f_{n}+ \text{H.c.}
\end{equation}
The odd fiber modes coupled to the mechanical modes in the sending node will be coupled to mechanical modes in the receiving node with the relative phase difference of $\pi$ due to the odd number of maxima in their intensity profile.
The phase factor $(-1)^{n}$ in the Hamiltonian dictates the phase difference at the two ends of the fiber for far-detuned odd modes.
For the sake of convenience, hereafter we denote the time-dependent coupling strengths $g_i(t)$ by $g_i$. The laser-enhanced time-dependent effective optomechanical coupling strength of the mechanical mode to the fiber mode is given by $g_{i} =\sqrt{\gamma_{e,i} \delta_{\textsc{fsr}}/2\pi}$ \cite{patel2018single,stannigel2011optomechanical}.
Here $\gamma_{e,i}\equiv G_i^2/\kappa$ is the rate at which the mechanical excitations are converted into the flying qubits in the fiber through the optical cavity. $G_i(t)$ is the time-dependent optomechanical coupling and $\kappa$ is the cavity decay rate.
Typically, $\gamma_{e,i}$ exceeds the intrinsic damping rate $\gamma_{m}$ of the mechanical resonators and thus $\gamma_{m}$ has negligible effects on the QST efficiency. The effect of finite mechanical damping rate is considered in Sec.~\ref{sec31}.

In the interaction picture of the Hamiltonian $\nonumber  H_{0}  = \hbar \omega _{m} (m_{1}^{\dag } m_{1} +m_{2}^{\dag } m_{2}+\sum_{n}f_{n}^{\dag } f_{n})$,
the system dynamics is governed by
\begin{eqnarray}\label{interaction Hamiltonian}
\tilde{H}&=&\hbar\! \sum_{n=-\infty}^{+\infty}(\Delta_{0}-\omega _{m}+n\delta_{\textsc{fsr}}) f_{n}^{\dagger }f_{n} \nonumber\\
&+&\hbar \sum_{n=-\infty}^{+\infty} \left[  g_{1}m_{1}^{\dag } +(-1)^{n} g_{2} m_{2}^{\dag } \right]f_{n}+ \text{H.c,}
\end{eqnarray}
where $m_{i=1,2}$ and $\omega_{m}$ are the annihilation operators and the frequency of the mechanical modes at the nodes, respectively. For the case in which a quantum state is transferred between the two mechanical modes, the pump lasers should be at the first red sideband. We, thus, take $\Delta_{0}=\omega _{m}$) as the working point from now on~\cite{wang2012d,felicetti2017quantum,dong2014optomechanically}.
We begin the analysis by writing the Heisenberg equations of motion in the interaction picture~\cite{walls2007quantum}:
\begin{equation}
\dot{V}(t)=-iM(t)V(t),
\label{heisenberg}
\end{equation}
where $V$ is the vector of mode operators and $M$ is the matrix of dynamics.
For a single-mode fiber one has $V = (m_1, f_0, m_2)^\intercal$ and
\begin{equation}\label{eq+2}
M(t) =
\left(\begin{array}{ccc}
		{0} & {g_{1} } & {0} \\
		{g_{1} } & {-\frac{i}{2}\gamma_{\rm fib}} & {g_{2} }\\
		{0} & {g_{2} } & {0 }
\end{array}\right).
\end{equation}
Generalization to the multi-mode case will be done later in Sec. \ref{sec32}.

\subsection{Transfer protocols\label{sec22}}
We first describe the formulation of two protocols used in this work for the deterministic QST between two mechanical resonators over an optical waveguide. One of the universal protocols for deterministic QST is the adiabatic passage. In the AP protocol, the classical driving lasers of both nodes are applied in a counterintuitive order, i.e., the receiving node, here $g_{2}$, precedes that of the sending node ($g_{1}$).
By introducing the mixing angle $\vartheta \equiv \tan^{-1}(g_{1}/g_{2})$ the above requirement translates to $\lim_{t\rightarrow t_{i}}(g_{1}/g_{2})=0$ and  $\lim_{t\rightarrow t_{f}}(g_{1}/g_{2})= \infty$, which implies that $\vartheta(t_{i})=0$ and $\vartheta(t_{f})=\pi/2$ \cite{vitanov1997analytic}.
In analogy to the dark and bright states in atom systems, in our effective three-mode optomechanical system the dark mode $A_{0}$ and two bright superposition modes $A_{\pm}$ are obtained as the following
\begin{subequations}\label{eq5}
\begin{align}
A_{+} &=\frac{1}{\sqrt{2}}\big( m_{1}\sin{\vartheta(t)} + m_{2}\cos{\vartheta(t)}+f_{0} \big), \\
A_{0} &=m_{1} \cos{\vartheta(t)} -m_{2} \sin{\vartheta(t)}, \\
A_{-} &=\frac{1}{\sqrt{2}}( m_{1}\sin{\vartheta(t)} + m_{2}\cos{\vartheta(t)}-f_{0}).
\end{align}
\end{subequations}
The adiabatic modes described in above relations are the instantaneous eigenmodes of the dynamic matrix in  Eq.~\eqref{eq+2} in the absence of fiber decay.
The AP mechanism is then easily understood by looking at the Eqs.~\eqref{eq5}. By preparing the three-mode system in the sending mechanical mode $m_{1}$ at the beginning of the transfer process, the goal is to transfer its quantum state to the receiving mode $m_2$ only through the adiabatic mode $A_{0}$.
The AP protocol employs large operation times and/or large driving pulses amplitudes to maintain the system always in the dark mode $A_{0}$. Hence, safely transferring the quantum state from one mode to the other without exciting the fiber modes. 

Generally, if the process duration $T$ or driving pulses strength is large enough, the system will evolve along with either of its adiabatic modes $A_{i}$ without transition among them.
This property reveals the weakness of the AP protocol; which is its slowness. It, therefore, can suffer from dissipations in the sending and receiving nodes as well as the loss in the channel during the entire procedure.
Nevertheless, if the process is fast, i.e. $T \ngtr T_{0}$, where $T_{0}$ is the characteristic time of the adiabatic passage determined by the system parameters, the time evolution path does not follow the adiabatic eigenmodes $A_{i}$ due to the diabatic transitions among them.
Therefore, it is vital to propose a technique to speed up the adiabatic passage and yet maintain the high efficiency. Indeed, one should remove the diabatic transitions among the adiabatic eigenmodes during the fast transfer process.
According to the transitionless quantum driving algorithm~\cite{berry2009transitionless,chen2010shortcut}, we introduce a modification to the dynamic matrix $M(t)$ in Eq.~\eqref{heisenberg} such that the non-adiabatic transitions are eliminated. This allows us to perform a rapid and efficient QST between the mechanical resonators in the system even with weak driving pulses. This, indeed, is done by adding an auxiliary counter-diabatic process to the system whose contribution to the dynamics is described by the following dynamical matrix
\begin{equation}\label{eq+1}
M_{\rm cd}(t)=i \sum_{k}\dot{A}^{}_{k}A_{k}^{\dagger},
\end{equation}
where the summation is over all adiabatic modes and the dot represents time derivation.
Such processes prohibits the inter-mode transitions among the system adiabatic eigenmodes.
By substitution from Eqs.~\eqref{eq5} in \eqref{eq+1} the counter-diabatic dynamical matrix $M_{\rm cd}$ for the three-mode system is obtained as the following
\begin{equation} \label{eq+6}
M_{\rm cd}= \left(\begin{array}{ccc}  {0} & {0} & {ig_{a}} \\ {0} & {0} & {0} \\ {-ig_{a}} & {0} & {0}  \end{array}\right).
\end{equation}
Here, $g_{a}=(\dot{g_{1}}g_{2}-g_{1}\dot{g_{2}})/g_{0}^{2}$ with $g_{0}=\sqrt{g_{1}^{2}+g_{2}^{2}}$ is the coupling rate to the auxiliary driving field.

To realize this scheme, one could employ an extra fiber mode that couples to both mechanical modes at the sending and receiving nodes.
However, we focus on the realization of the counter-diabatic driving in our system without additional coupling between the two mechanical resonators.
Specifically, we develop a feasible approach to implement STAP protocol.
It is straight forward to show that $M_{\rm cd}(t)$ can be absorbed into the variation of the reference pulses.
In other words, no additional driving field is required to achieve $M_{\rm cd}(t)$.
The dynamic of our system in the absence of fiber losses satisfies SU(3) Lie algebra, which simplifies the shortcut to adiabatic passage design.
In order to cancel the counter-diabetic processes yet maintaining the same dynamics, we employ the unitary transformation
\begin{equation} \label{eq+10}
U(t)= \left(
\begin{array}{ccc}
	{1} & {0} & {0} \\
	{0} & {\cos~\phi(t)} & {i\sin~\phi(t)} \\
	{0} & {i\sin~\phi(t)} & {\cos~\phi(t)}
\end{array}\right).
\end{equation}
With this transformation the total dynamic matrix in the absence of fiber losses becomes
\begin{equation} \label{eq+11}
\widetilde{M}_{t}(t)= \tilde{g}_{1}\mathcal{G}_{1}+ \tilde{g}_{2}\mathcal{G}_{6}-\tilde{g}_{a}\mathcal{G}_{5},
\end{equation}
where $\mathcal{G}_{i}~(i=1,2,6)$ are Gell-Mann matrices and 
\begin{subequations}\label{eq+12}
	\begin{align}
	\tilde{g}_{1} &={g}_{1}\cos\phi-{g}_{a}\sin\phi, \\
	\tilde{g}_{2} &={g}_{2}+\dot\phi, \\
	\tilde{g}_{a} &={g}_{1}\sin\phi+{g}_{a}\cos\phi.
	\end{align}
\end{subequations}
Setting $\tilde{g}_{a}=0$ gives $\phi=\arctan(-{g}_{a}/{g}_{1})$ and
\begin{subequations}\label{eq+13}
\begin{align}
\tilde{g}_{1} &=\sqrt{{g}_{1}^{2}+{g}_{a}^{2}}, \\
\tilde{g}_{2} &={g}_{2}+\dot \phi.
\end{align}
\end{subequations}
Therefore, the STAP protocol can be realized by replacing the reference driving pulses ${g}_{1}$ and ${g}_{2}$ in dynamic matrix \eqref{eq+2} with the modified driving pulses $\tilde{g}_{1}$ and $\tilde{g}_{2}$, respectively.
In the following, we study the STAP protocol and show that it allows a fast and efficient state transfer between two mechanical modes by only reshaping the original driving pulses. More interestingly, the driving pulses do not need to satisfy the adiabaticity conditions such as large duration time and strength of the driving pulses.

\section{\label{sec3}results and discussion}
\begin{figure}[t]
\includegraphics[width=0.7\columnwidth]{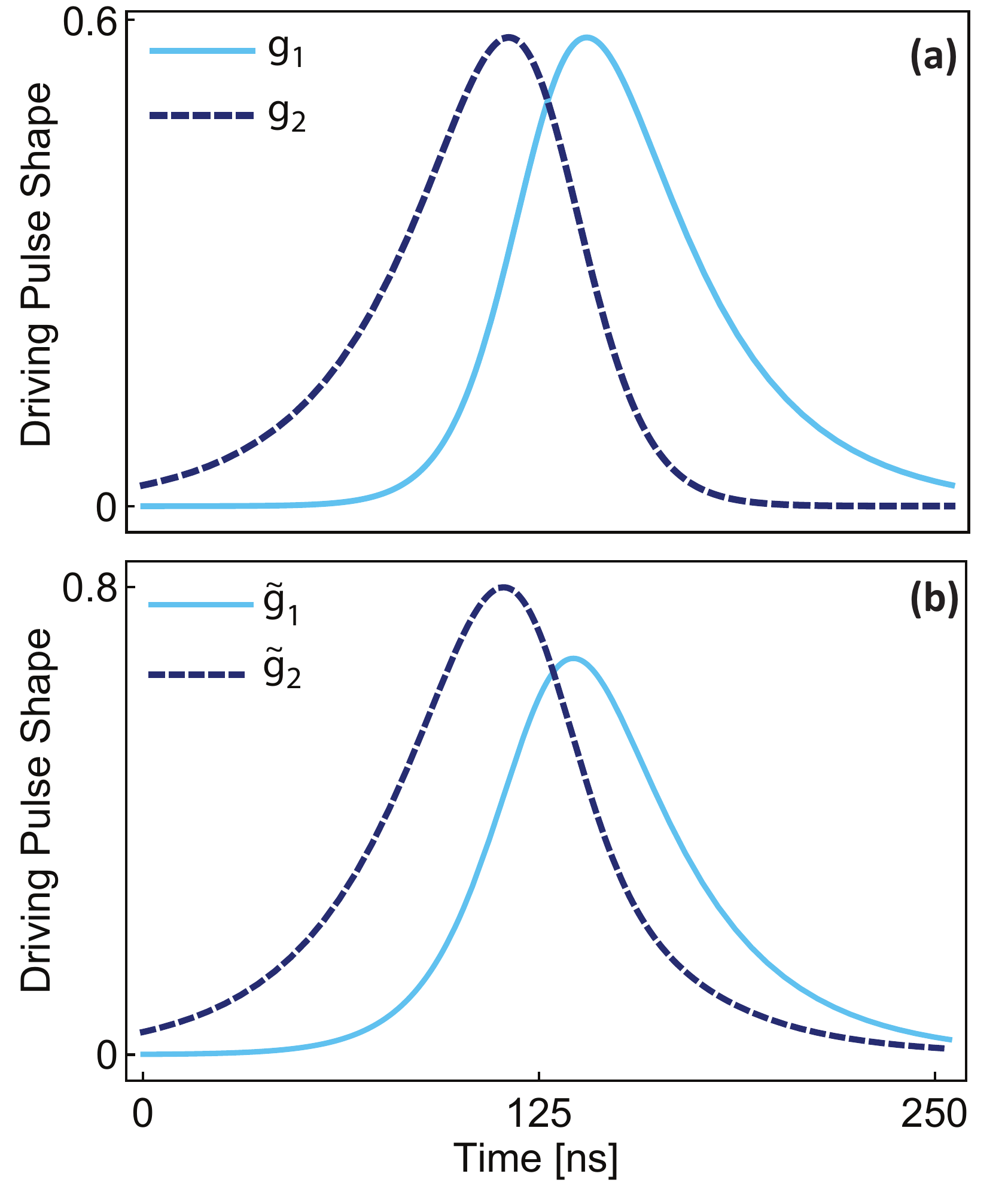}
\caption{(a) The original (\ref{eq4}) and (b) modified (\ref{eq+13}) pulse shapes used in protocols AP and STAP. The parameters are $\lambda_{0}/2\pi = 10$ $\rm MHz$, $T=250 $ $ \rm ns$, $\sigma=T/8$.}
\label{fig2}
\end{figure}
We evaluate the performance of above mentioned deterministic QST protocols between two mechanical resonators over an optical fiber.
We define the QST efficiency $\epsilon$ as the square of the ratio of final state of mechanical mode in the second node $m_{2}(t_{f})$ to initial state of mechanical mode in the first node $m_{1}(t_{i})$.
For performing the adiabatic state transfer, we consider the time-dependent coupling functions strength as an adiabatic reference,
\begin{subequations}
\label{eq4}
\begin{align}
g_{1}(t) &=g_{0}(t) \sin[\frac{\pi}{4} s(t)], \\
g_{2}(t) &=g_{0}(t) \cos[\frac{\pi}{4} s(t)],
\end{align}
\end{subequations}
where $g_{0}(t)=\frac{1}{\sqrt{2}} ~\lambda_{0}~\text{sech}[\frac{1}{\sigma}(t-T/2)]$, $s(t)=1+\tanh[\frac{1}{\sigma}(t-T/2)]$ with $T$ the coupling time duration and $\sigma$ the semi-width at half-maximum of the coupling function strength. Fig.~\ref{fig2} shows the original (\ref{eq4}) and modified (\ref{eq+13}) pulse shape that we apply in the AP and STAP protocols, respectively.

\subsection{\label{sec31} Single-mode fiber}
\begin{figure}[b]
\includegraphics[width=0.7\columnwidth]{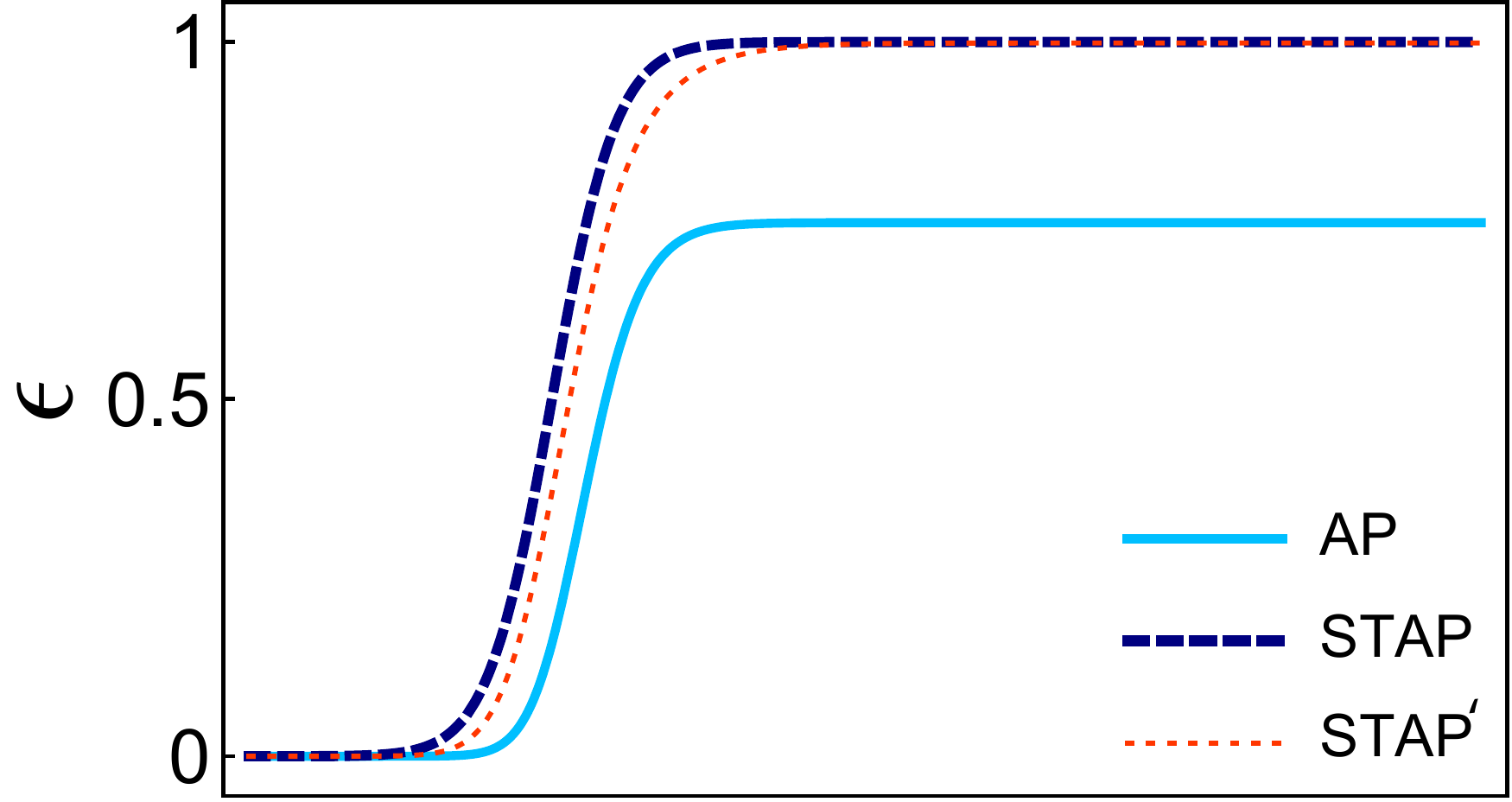}
\includegraphics[width=0.7\columnwidth]{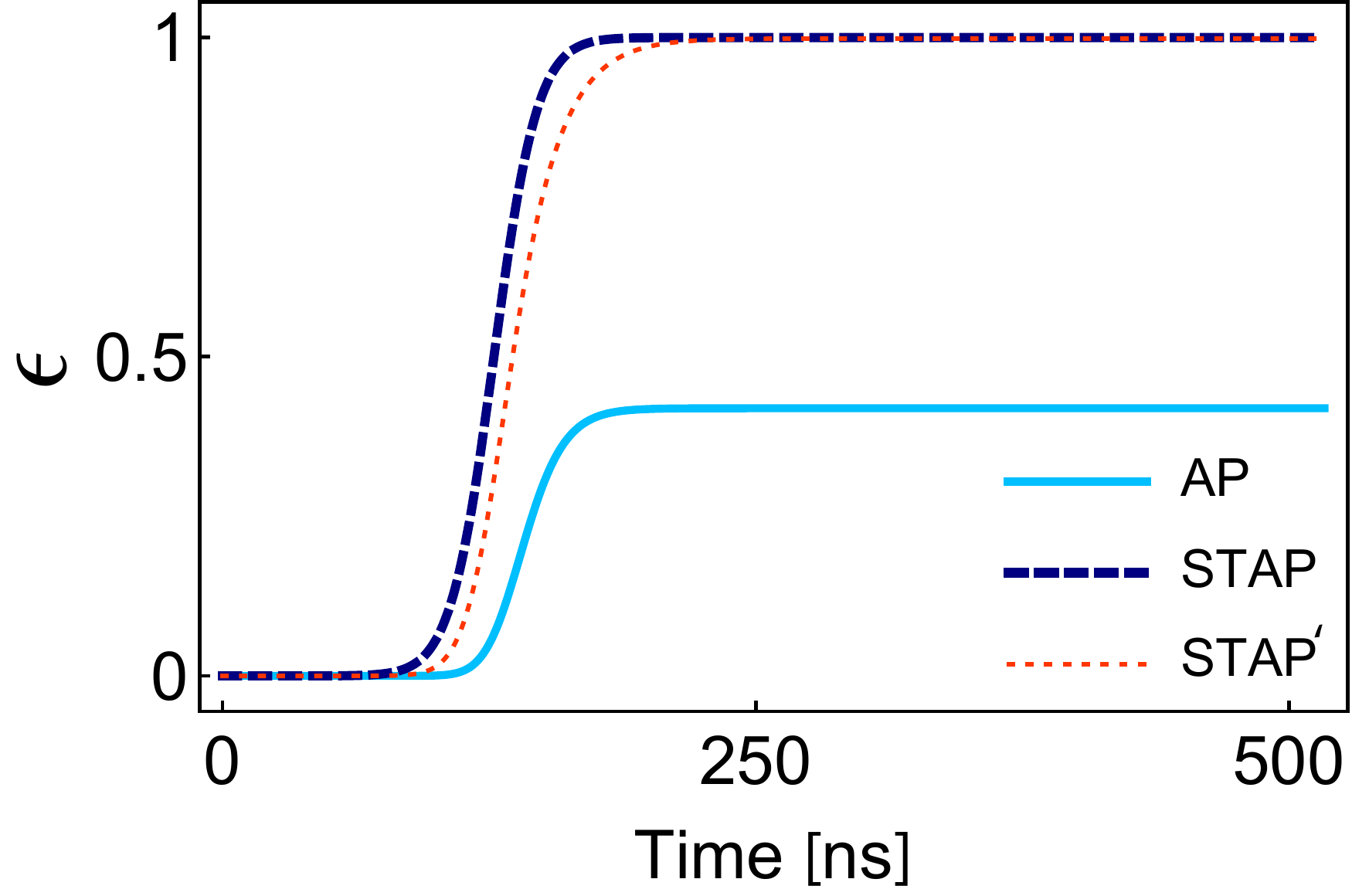}
\caption{The efficiency of QST as a function of the operation time for adiabatic passage (solid blue line), shortcut to adiabatic passage (dashed black line), and effective STAP (dotted red line): (Top) $\sigma=T/6$, (Bottom) $\sigma=T/8$. The other parameters are $\gamma_{\rm fib}/2\pi = 22 $~kHz, $\lambda_{0}/2\pi = 10$~MHz, and $T=250$~ns.}
\label{fig3}
\end{figure}
\begin{figure}[h]
\includegraphics[width=0.7\columnwidth]{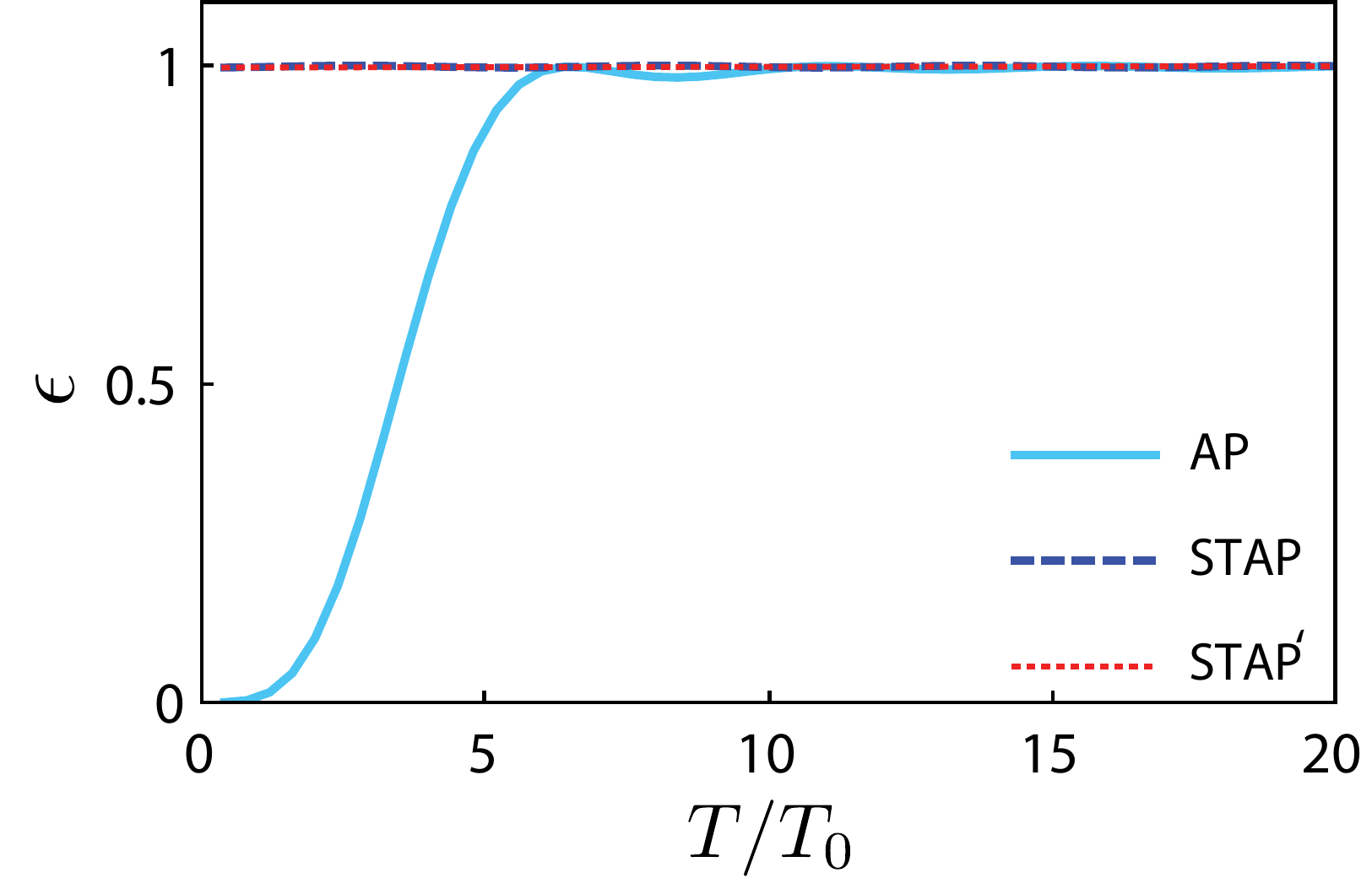}
\caption{The efficiency of QST as a function of the operation time for both protocols. $\sigma=T/8$ while the other parameters are as Fig.~\ref{fig3}.}
\label{fig4}
\end{figure}
To describe the physics behind the above mentioned protocols, first, we consider the single-mode fiber instead of multi-mode fiber. In Fig.~\ref{fig3}, we study the effect of the coupling functions time duration that leads to accelerating the QST process on the protocols efficiency. It is found that in the AP protocol as the speed of the QST process increases, i.e. the coupling functions time duration is smaller, the transfer efficiency decreases. Because the speeding up transfer process in the AP protocol leads to the system evolution does not follow the dark mode and to have transitions between other adiabatic modes. However, in STAP protocol the transfer efficiency becomes unity for any coupling function time duration since the non-adiabatic transition has been removed by adding the term $M_{\rm cd}$ [see Fig.~\ref{fig4}].
Also, in Fig.~\ref{fig3} it can be seen that the considered time duration is not sufficient for a perfect QST in the AP protocol.
As mentioned before for a complete state transfer between two mechanical modes we need to increase the time in the AP protocol.
However, we notice that at the time duration which the adiabatic condition is not fulfilled, one still achieves an efficient state transfer by the STAP protocol.

To quantitatively study the efficiency, we plot the transfer efficiency versus the time duration $T$ in Fig.~\ref{fig4}. It is obvious from the figure that in the AP protocol, the time duration $T$ required to reach perfect transfer should be much longer than the characteristic time of the adiabatic passage $T_{0}=\sqrt{2}\pi/\lambda_{0}$ ($T \gg T_{0}$), where the adiabatic condition is fulfilled and the system follows the dark mode.
However, we observe that in the STAP protocol the transfer efficiency becomes unity for any time duration $T$ even for short durations. Therefore, the protocol offers a fast and efficient QST between mechanical modes. The red dotted curve in Figs. \ref{fig3} and \ref{fig4} has been plotted by replacing the reference driving pulses ${g}_{1}$ and ${g}_{2}$ in dynamic matrix \eqref{eq+2} with the modified driving pulses $\tilde{g}_{1}$ and $\tilde{g}_{2}$. It fits well with the results of adding an auxiliary counter-diabatic term (the dark blue dashed curve) and verifies that a proper modification in the adiabatic reference pulses is equivalent to introducing a counter-adiabatic process.

\begin{figure}[b]
	\includegraphics[scale=.55]{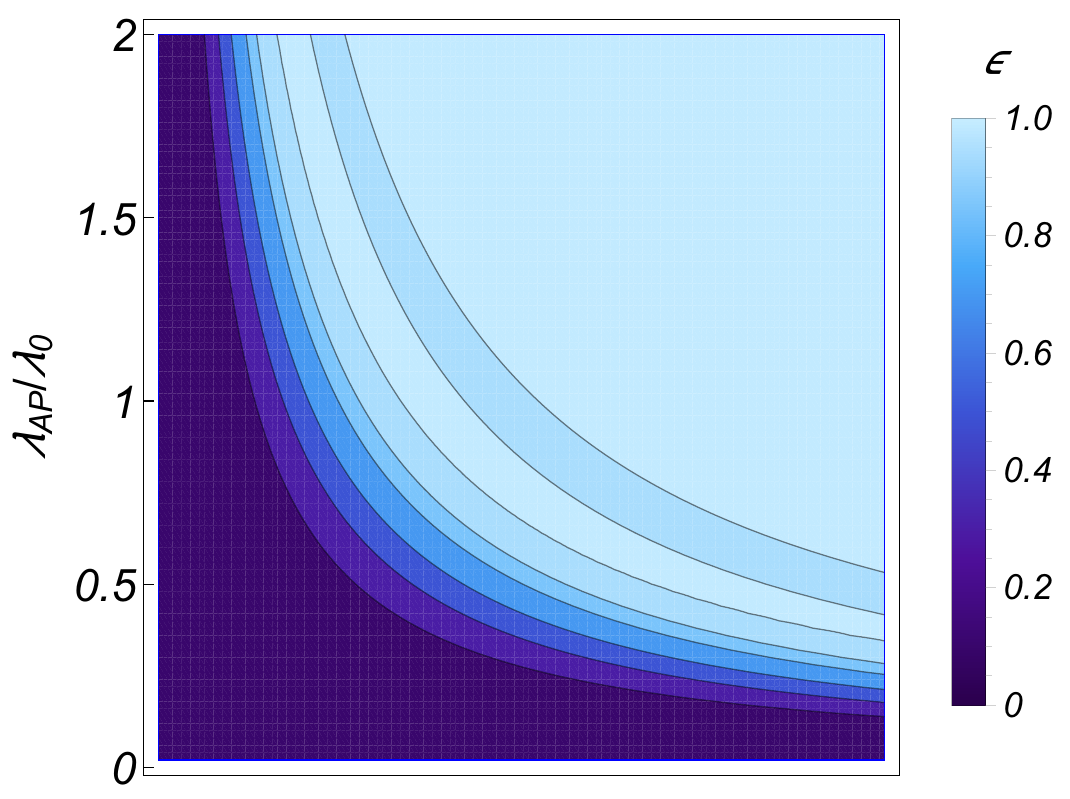}
	\includegraphics[scale=.55]{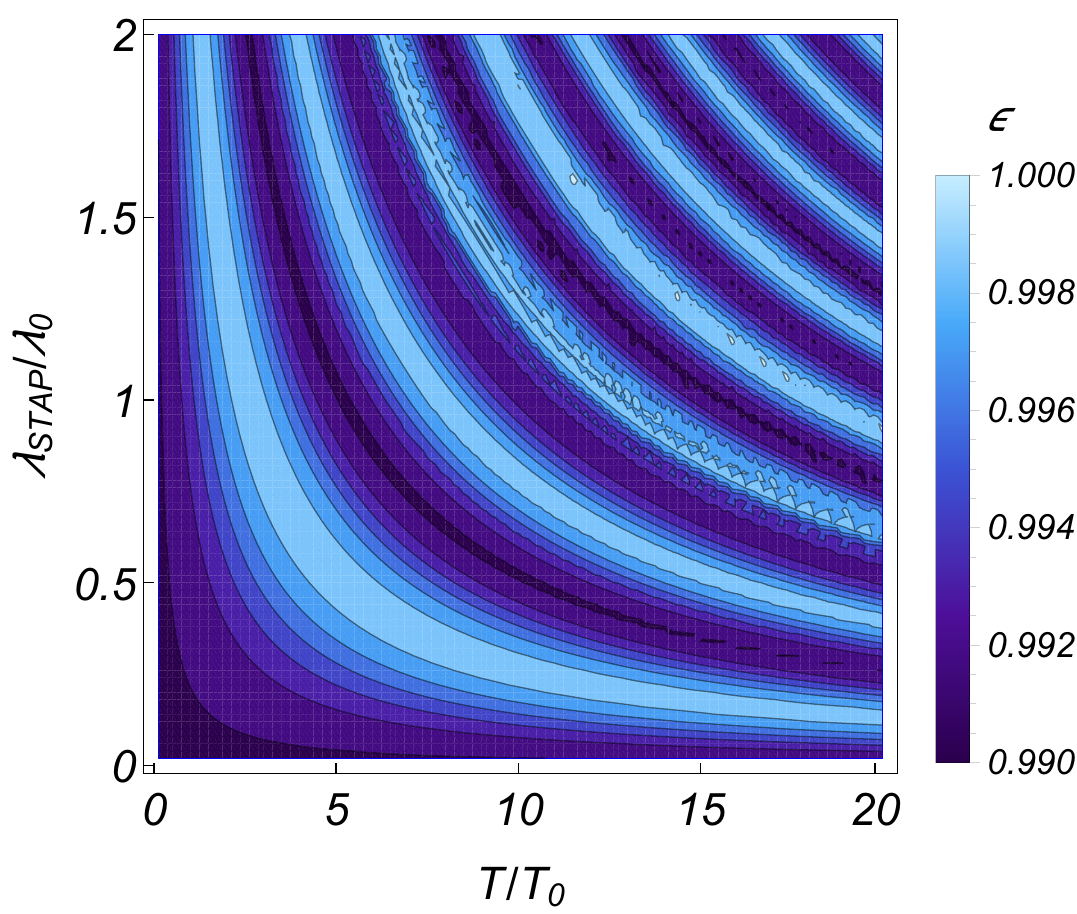}
	\caption{The efficiency of QST as a function of pulse duration and the maximum coupling strength for the (Top) AP protocol, (Bottom) STAP protocol.
		The parameters are the same as Fig.~\ref{fig4}.}
	\label{fig5}
\end{figure}

To illustrate some of the STAP protocol superior features to the AP protocol, we plot the transfer efficiency versus the peak of driving pulses strength and time duration for both protocols in Fig.~\ref{fig5}.
The maximum driving pulses strength required for the transfer efficiency to become close to unity in the STAP protocol is much less than that of the AP protocol. It is found that transfer efficiency becomes unity for a wide range of the strength of the driving pulse in the STAP protocol while it is unity for a small range of the large driving pulses strength in the AP protocol.
Therefore, in contrast to the AP process, the transfer efficiency in the STAP protocol is not dramatically affected by any parameters in the coherent driving fields.
Fig.~\ref{fig5} shows that for achieving a nearly perfect state transfer in a given time, the required increase in the strength of the drive pulses is much less in the STAP in comparison to the AP protocol.
Thus, by using the STAP protocol, attains a fast and efficient QST even with a very small value of the coupling strength.

We now analyze behavior of the QST efficiency for the two protocols in the presence of resonators damping. First we remind that a better adiabaticity is acquired by increasing the time duration. On the other hand, in state transfer process the time duration should be shorter than the mechanical damping time scales. For this reason, a highly efficient QST may be impossible for so large time duration in the presence of resonators damping. The conflict between transfer efficiency and speed results-in an optimal intermediate transferring time $T_{\rm op}$.
It is found that the efficient QST is possible when $T_{0}\ll T_{\rm op} \ll 1/ \gamma_{m},1/ \gamma_{\rm fib}$.
In this regime the QST can be both reasonably rapid compared to the resonators damping rate $\gamma_{m}$ and yet adiabatic enough to prevent excitation to the bright states.
As a result the resonators damping has negligible effects on the QST.
The STAP protocol is a fast process and therefore does not suffer significantly from the resonators damping in the sending and receiving nodes.
The STAP protocol is an appropriate protocol when both speed and the dissipation of nodes are problematic.

\subsection{\label{sec32} Multi-mode fiber}
In the following, we investigate the effect of higher order modes
 of the fiber on the transfer efficiency in the adiabatic passage.
To begin, we restrict our calculations to the fiber central mode $f_{0}$ and the first pair of neighboring modes $f_{\pm 1}$.
This brings us to a set of Heisenberg equations of motion as in Eq.~\eqref{heisenberg} with the vector operator $V(t)=(m_{1}, f_{-1}, f_{0}, f_{+1}, m_{2})^\intercal$ and the dynamical matrix
\begin{equation*}\label{dynamic matrix 1}
M(t)=
\left(\begin{array}{ccccc}
	{0} & {g_{1}} & {g_{1}} &{g_{1}} &{0} \\
	{g_{1}} & {-\delta_{\textsc{fsr}}-\frac{i}{2}\gamma_{\rm fib}} & {0}& {0} & {-g_{2}} \\
	{g_{1}} & {0} & {-\frac{i}{2}\gamma_{\rm fib}} & {0} & {g_{2}} \\
	{g_{1}} & {0} & {0} & {\delta_{\textsc{fsr}}-\frac{i}{2}\gamma_{\rm fib}} & {-g_{2}} \\
	{0} & {-g_{2}} & {g_{2}} &{-g_{2}} &{0}
\end{array}\right).
\end{equation*}
\begin{figure}[t]
\includegraphics[width=0.7\columnwidth]{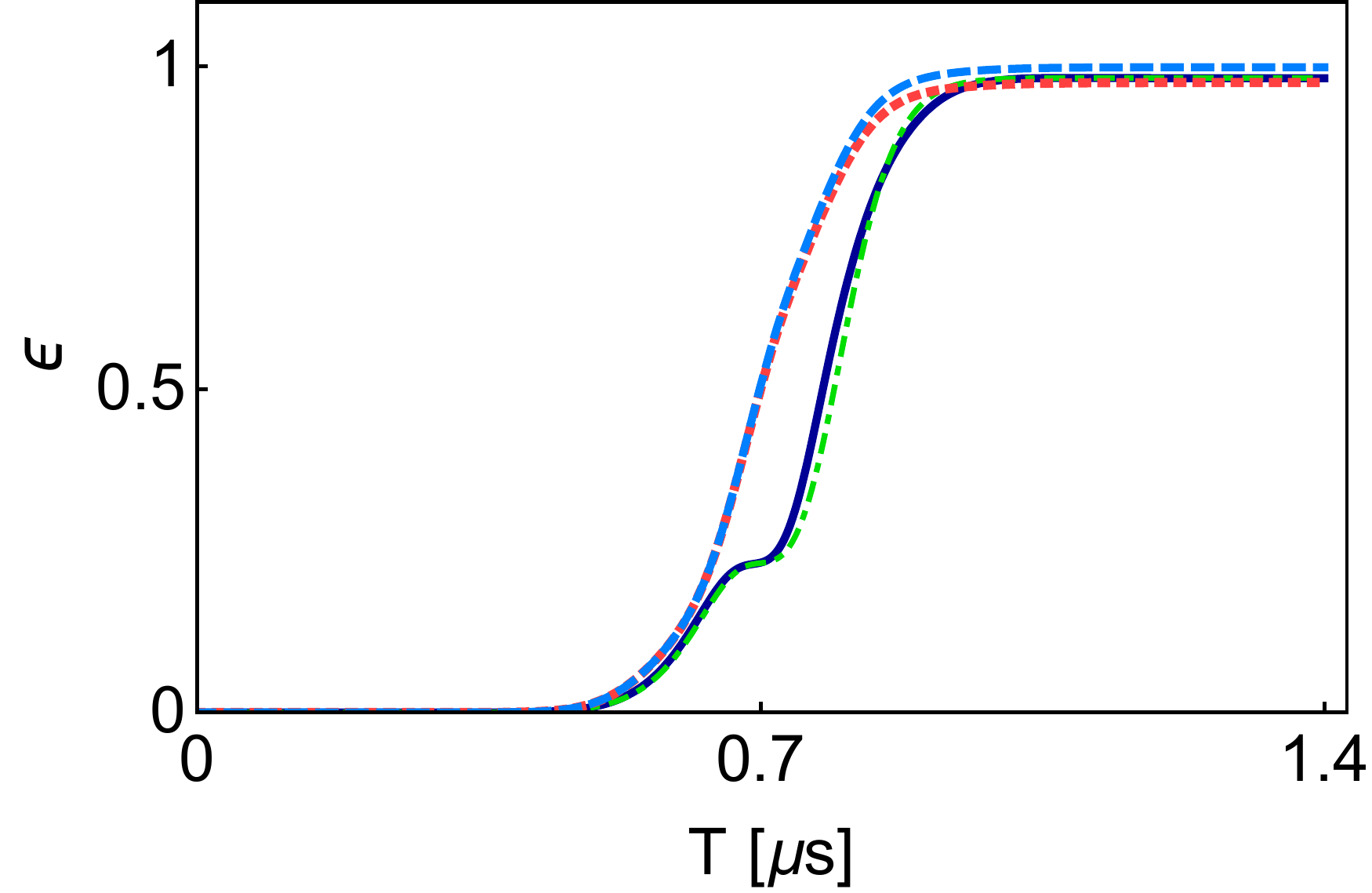}
\caption{The efficiency of QST as a function of the time duration for single-mode fiber (blue dashed line), three-mode fiber (dark solid line) and nineteen-mode fiber (green dash-dotted line).
The red dotted curve shows the result of numerical integration of Eqs.~\eqref{dynamic matrix}. Here, $\delta_{\textsc{FSR}}/2\pi = 10$~MHz and the other parameters are the same as Fig.~\ref{fig4}.
\label{fig6}}
\end{figure}
In writing the above matrix, we have taken into account the $\pi$-phase difference in the coupling of the odd modes $f_{\pm1}$ to the mechanical modes in the receiving node.
Zero-eigenvalue eigenmode of the above dynamics matrix for three-mode fiber in the absence of fiber decay reads
\begin{equation} \label{zero-eigenvalue}
A_{0} = \mathcal{N}
\Big({g_{2}}, {\frac{2g_{1}g_{2}}{\delta_{\textsc{fsr}}}}, 0, {-\frac{2g_{1}g_{2}}{\delta_{\textsc{fsr}}}}, {-g_{1}}\Big)^{\intercal},
\end{equation}
where $\mathcal{N}$ is the normalization coefficient.
Even though the superposition mode $A_{0}$ is dark with respect to the central fiber mode $f_{0}$, it still has a finite coupling to the first odd modes $f_{\pm1}$ and hence these modes can get populated during the QST process.
In other word, the adiabatic mode $A_{0}$ is not completely dark when other modes of the fiber are taken into account.
To investigate the effect of this leakage of the information we perform numerical calculations. The results are presented in Fig.~\ref{fig6} and show that the QST remains efficient for three-mode fiber in the adiabatic limit where the time duration is $20~T_{0}\approx 1.4 $ $ \rm \mu s$, see dark blue solid curve.
Therefore, a more general adiabatic-transfer (AT) mode exists which links the mechanical modes $m_{1}$ and $m_{2}$.
Obviously, from zero-eigenvalue eigenmode $A_{0}$ can be seen that the amplitude of odd modes $f_{\pm1}$ are nonzero during the transfer process but vanish at the beginning and end of the adiabatic passage protocol, because $g_{1}$ and $g_{2}$ vanish at $t=0$ and $t=T$, respectively.
Hence, the zero-eigenvalue adiabatic mode $A_{0}$ in Eq.~(\ref{zero-eigenvalue}) is an AT mode in the adiabatic limit and effectively resembles an efficient dark mode.

In order to generalize this conclusion and a more complete description of the problem, we include more fiber mode $f_{n}$ with $n \in[-9,+9]$ and solve the problem numerically.
The numerical integration of the Heisenberg equations are presented in Fig.~\ref{fig6} with the green dot-dashed line.
It is obvious that even for nineteen-mode fiber an AT mode exists that efficiently transfers quantum states between the two mechanical modes in the adiabatic limit.
For large time duration where the adiabatic condition is fulfilled, the effective AT mode performs as a perfect dark mode.
During the process and at short times duration due to the involvement of transient contributions from multi-mode fiber odd modes in the AT mode, the QST efficiency through AT mode is smaller than that of going through a perfect dark mode. 

To further illustrate that far-detuned modes of the long fiber $f_{\pm n}$ do not affect the transfer efficiency in the adiabatic limit, we study the effective single-mode model and derive an analytical expression for the transfer efficiency in the adiabatic limit.
We begin by solving Eq.~\eqref{heisenberg} for a three-mode fiber.
By assuming $\frac{1}{4}\gamma_{\rm fib}^{2} +\delta_{\textsc{fsr}}^{2} \gg g_{1(2)}^{2} $ one adiabatically eliminates the $f_{\pm 1}$ modes and reach to the following effective set of equations
\begin{equation} \label{dynamic matrix}
i\left(\begin{array}{ccc}  {\dot m_{1}}\\{\dot f_{0}}\\{\dot m_{2}}  \end{array}\right)=\left(\begin{array}{ccc} {-i\Gamma_{1}} & {g_{1} } & {i\Gamma _{12} } \\ {g_{1} } & {-\frac{i}{2}\gamma_{\rm fib}} & {g_{2} } \\ {i\Gamma _{12}} & {g_{2} } & {-i\Gamma _{2} }  \end{array}\right)  \left(\begin{array}{ccc}  {m_{1}}\\{f_{0}}\\{m_{2}}  \end{array}\right),
\end{equation}
where we have introduced $\Gamma_{12} =g_{1} g_{2} \gamma_{\rm fib}(\frac{1}{4}\gamma_{\rm fib}^{2} +\delta_{\textsc{fsr}}^{2})^{-1} $ and $\Gamma _{i} =g_{i}^{2} \gamma_{\rm fib}(\frac{1}{4}\gamma_{\rm fib}^{2} +\delta_{\textsc{fsr}}^{2})^{-1}$ with $(i=1,2)$.
Indeed, when the $\delta_{\textsc{fsr}}$ is larger than the cavity decay rate ($\delta_{\textsc{fsr}} \gg \kappa$) the mechanical resonators effectively only couple to the central fiber mode. Therefore, assuming an effective single-mode fiber is a valid consideration.
The red dotted curve in Fig.~\ref{fig6} verifies our anticipation as it shows the result of the numerical integration of Eqs.~\eqref{dynamic matrix} and matches well with the results of the single-mode fiber case, the blue dashed curve.

In the adiabatic representation, the Eq.~\eqref{dynamic matrix} turns into

\begin{widetext}
\begin{equation}\label{adia}
\left(\begin{array}{ccc}{\dot A_{+}}\\{\dot A_{0}}\\{\dot A_{-}}  \end{array}\right)
= \Bigg\lbrace   \left(\begin{array}{ccc} {-ig_{0}^{2}(t) -\frac{1}{4}\gamma_{\rm fib} } & {\frac{1}{\sqrt{2}}\dot{\vartheta}} & {\frac{1}{4}\gamma_{\rm fib}} \\
{-\frac{1}{\sqrt{2}}\dot{\vartheta}} & {0 } & {-\frac{1}{\sqrt{2}}\dot{\vartheta}} \\
{\frac{1}{4}\gamma_{\rm fib}} &  {\frac{1}{\sqrt{2}}\dot{\vartheta}} &{ig_{0}^{2}(t) -\frac{1}{4}\gamma_{\rm fib} }
\end{array}\right)
-\eta  \left(\begin{array}{ccc} {\cos^{2}(\pi s)} &{-\frac{1}{\sqrt{2}}\sin(2\pi s)} & {\cos^{2}(\pi s)} \\ {-\frac{1}{\sqrt{2}}\sin(2\pi s)} & {2\sin^{2}(\pi s)} & {-\frac{1}{\sqrt{2}}\sin(2\pi s)}\\{\cos^{2}(\pi s)} & {-\frac{1}{\sqrt{2}}\sin(2\pi s)} &{\cos^{2}(\pi s)} \end{array}\right)\Bigg\rbrace     \left(\begin{array}{ccc}  {A_{+}}\\{A_{0}}\\{A_{-}}  \end{array}\right),
	\end{equation}
\end{widetext}
with $\eta=g_{0}^{2}(t) \gamma_{\rm fib}(\frac{1}{2}\gamma_{\rm fib}^{2} +2\delta_{\textsc{fsr}}^{2})^{-1}$.
\begin{figure}[t]
	\includegraphics[width=0.8\columnwidth]{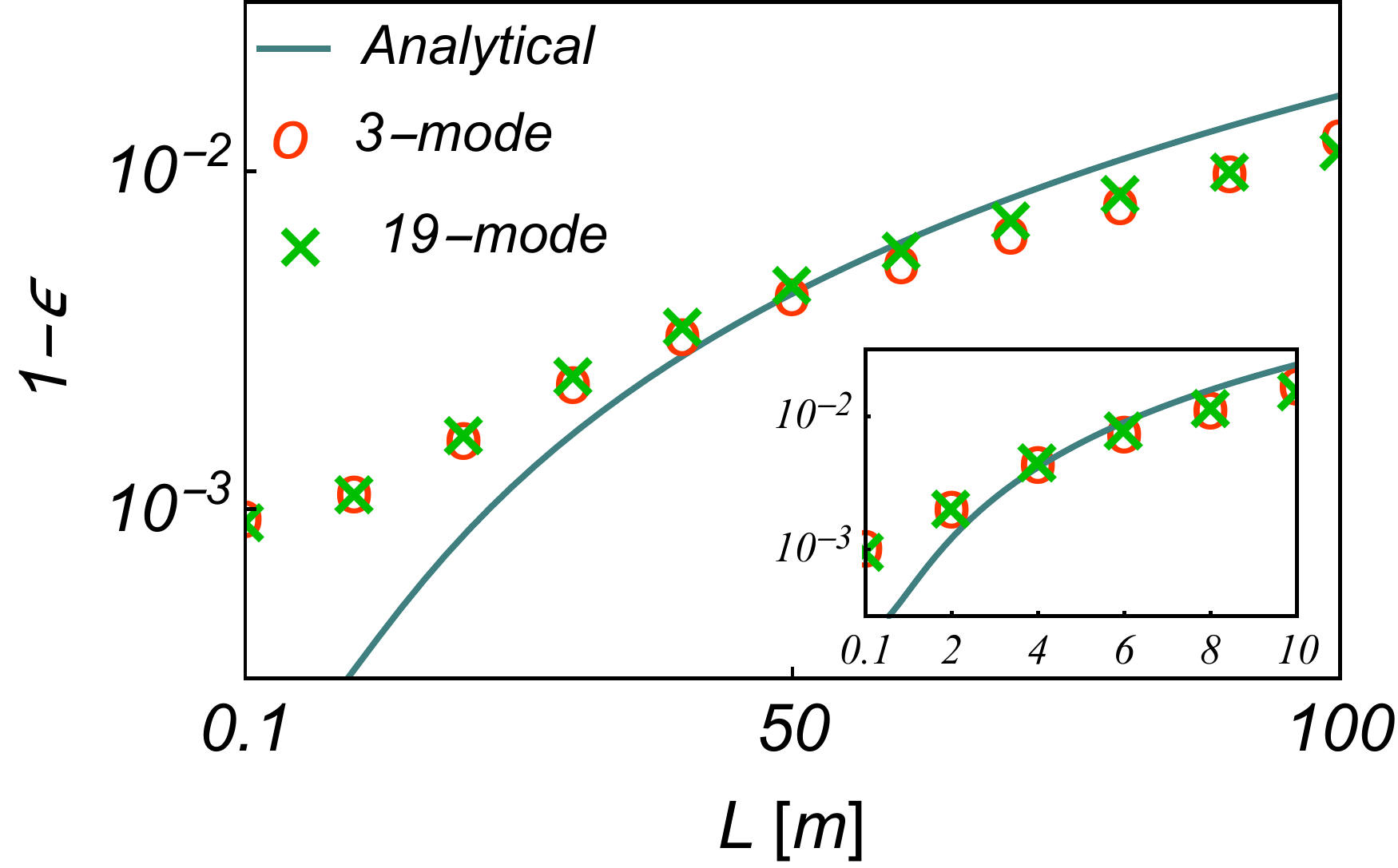}
	\caption{The inefficiency of QST as a function of fiber length in the adiabatic limit.
		The red circles are for the three-mode fiber, green crosses for the nineteen-mode fiber, while the solid curve is for the analytical expression in Eq.~\eqref{m2}.
		The parameters are $T=20T_{0}$, $T_{0}=\sqrt{2}\pi/\lambda_{0}$, $c=2\times 10^{8}$~m/s, and $\sigma=T/4$; (main figure) $\lambda_{0}/2\pi= 1$~MHz, $\gamma_{\rm fib}/2\pi= 1.5$~kHz, and (inset) $\lambda_{0}/2\pi= 10$~MHz, $\gamma_{\rm fib}/2\pi= 22$~kHz.}
	\label{fig7}
\end{figure}
When $\lambda_{0}$ is large compared to $\gamma_{\rm fib}$, the diagonal elements
in the equations for $A_{+}$ and $A_{-}$ in Eqs.~\eqref{adia} dominate over the off diagonal ones. Then we can carry out adiabatic elimination
of modes $A_{+}$ and $A_{-}$ by setting $\dot A_{+}=\dot A_{-}=0$ and eliminating $A_{+}$ and $A_{-}$ from the resulting set of two algebraic equations. Therefore, in the limit $\gamma_{\rm fib} \ll \lambda_{0}$, $\gamma_{\rm fib} \ll \delta_{\textsc{fsr}} $ and $\frac{1}{4}\gamma_{\rm fib}^{2} +\delta_{\textsc{fsr}}^{2} \gg g_{1(2)}^{2} $, we obtain
\begin{subequations}
\begin{align}
A_{+} &=\frac{(-ig_{0}+\gamma_{\rm fib}/4)(\dot{\vartheta} A_{0} / \sqrt{2})}{g_{0}^{2}}, \\
A_{-} &=\frac{(ig_{0}+\gamma_{\rm fib}/4)(\dot{\vartheta} A_{0} / \sqrt{2})}{g_{0}^{2}},
\end{align}
\end{subequations} 
and for $\sigma=T/8$ one arrives at
\begin{equation}
A_{0}(T)=A_{0}(0) \exp\Big\{\!-\frac{\gamma_{\rm fib} \pi^{2}}{2\lambda_0^{2}T}-\frac{\gamma_{\rm fib} \lambda_0^{2}T}{16\delta_{\textsc{fsr}}^{2}}\Big\}.
\end{equation}
Forasmuch as the dark mode $A_{0}$ equals to the mechanical modes $m_{1}$ and $m_{2}$ at the beginning and end of transfer process, respectively, we are brought to
\begin{equation} \label{m2}
\epsilon =\exp\Big\{\!-\frac{\gamma_{\rm fib} \pi^{2}}{\lambda_0^{2}T}-\frac{\gamma_{\rm fib} \lambda_0^{2}T}{8\delta_{\textsc{fsr}}^{2}}\Big\}.
\end{equation}
In Fig.~\ref{fig7} we compare this analytical expression for the efficiency with the numerical calculations. The curves present inefficiency $1-\epsilon$ of the protocol as a function of the fiber length.
The solid curve in Fig.~\ref{fig7} that displays analytical results based on Eq.~\eqref{m2} coincide with the numerical values obtained for three and also nineteen-mode fibers.
The plot suggests that an efficient QST with an error of about one percent across distances as long as hundreds of meters are practical, provided the chosen control parameters satisfy the adiabatic condition $T_{0}\ll T \ll 1/ \gamma_{\rm fib}$.
We emphasize that this is because of the existence of an AT mode in the multi-mode fibers which the QST is efficient through it in the adiabatic limit.

\section{\label{sec4}conclusion}
We have studied the application of two quantum state transfer protocols on the mechanical resonators coupled via a lossy optical fiber.
By the shortcut to the adiabatic passage protocol, one eliminates the diabatic transitions between adiabatic modes of the system during the transfer process.
We have shown that this leads to a fast and efficient state transferring through the dark mode evolution.
Meanwhile, the adiabatic passage protocol has a challenge in a conflict between the transfer speed and efficiency.
It was also shown that by the STAP protocol, the QST is possible even with very weak coupling pulses.
These remarkable features of the STAP protocol enable us to efficiently perform QST by the available resources and practical limitations, including mechanical dissipation.
Also, we have shown that for long-distance quantum communications, the adiabatic passage can mitigate the effect of transmission losses in a multi-mode fiber on transfer efficiency in the adiabatic limit.
An adiabatic transfer mode exists which transfers a quantum state between the mechanical modes with high efficiency in the adiabatic limit.
This study may pave the way for the use of optomechanical systems for the experimental realization of continuous-variable and long distances quantum communications.

\begin{acknowledgments}
The work of MR and KJ has been supported by Ferdowsi University of Mashhad.
MA acknowledges support by INSF (Grant No. 98005028).
\end{acknowledgments}

\bibliography{version1}

\end{document}